# CRNNTL: convolutional recurrent neural network and transfer learning for QSAR modelling

Yaqin Li,[1*] Yongjin Xu,[2] Yi Yu,[2*]

[1] West China Hospital, Sichuan University, 610041 Chengdu, PR, China
[2] Department of Chemistry and Molecular Biology, University of Gothenburg, Kemivägen 10, 41296 Gothenburg, Sweden

**ABSTRACT:** Molecular latent representations, derived from autoencoders (AEs), are widely used for drug or material discovery over past couple of years. In particular, a variety of machine learning methods based on latent representations has shown excellent performance on quantitative structure–activity relationship (QSAR) modelling. However, the sequence feature of them hasn't been considered in most cases. In addition, data scarcity is still the main obstacle for deep learning strategies, especially for biological activity datasets. In this study, we propose the convolutional recurrent neural network and transfer learning (CRNNTL) method inspired by the applications of polyphonic sound detection and electrocardiogram classification. Our strategy takes advantages of both convolutional and recurrent neural networks for feature extraction, as well as the data augmentation method. Herein, CRNNTL is evaluated on 20 benchmark datasets in comparison with baseline methods. In addition, one isomers-based dataset is used to elucidate its ability for both local and global feature extraction. Then, knowledge transfer performance of CRNNTL is tested, especially for small biological activity datasets. Finally, different latent representations from other type of AEs were used for versatility study of our model. The results show the effectiveness of CRNNTL using different latent representation. Moreover, efficient knowledge transfer is achieved to overcome data scarcity considering binding site similarity between different targets.

## Introduction

For the excavation of crucial molecular factors on properties and activities, quantitative structure–activity relationship (QSAR) has been an active research area in the past 50+ years.[1,2] In QSAR, the molecular representations (or descriptors), as the input features of the modelling, represent chemical information of actual entities in a computer-understandable numbers. Historically, molecular fingerprints, like extended-connectivity fingerprints (ECFPs) have been widely used as representations for the modelling of physicochemical, physiological properties and biological activities.[3] Recent development in deep neural network facilitates the motivation for the utilization of different molecular representations such as latent representations.[4,5] Latent representations are fixed-length continuous vectors which is derived from autoencoders (AEs) with encoder–decoder architecture.[6] Although AEs is first shown as a generative algorithm for de novo design studies,[7,8] latent representations from encoders of AEs have been extracted for QSAR modelling.[5] Without sophisticated human-engineered feature selection, these kinds of representations show competitive performance compared with traditional ones.[4] Furthermore, the representations and QSAR models can be used for multi-objective molecular optimization to tackle inverse QSAR problem.[9]

To obtain latent representations, sequenced-based strings, like SMILES (Simplified Molecular Input Line Entry Specification), are employed as inputs into AEs. The method in detail is described in the next section. In recent years, great interest has been aroused by developing QSAR models based on latent representations. One of the earliest developments in the field was the chemical variational autoencoders (VAE) by Aspuru-Guzik et al.[6] After generating latent representations for the encoder, two fully connected artificial neural network (ANN) were used for the prediction of water–octanol partition coefficient (logP), the synthetic accessibility score, and drug-likeness. More recently, Winter et al. has shown that latent representation applied on support vector machine (SVM) outperforms ECFPs and graph-convolution method by the translation AE model (named CDDD).[5] Subsequently, convolutional neural network (CNN) was trained for the QSAR prediction after generating latent representations.[10,11] Owing to the architectural characteristic of local connectivity, shared weight, and pooling,[12] some literatures indicated that CNNs perform better in QSAR modelling, compared with ANN and other traditional models.[13-16]

On the one hand, latent representations should have sequence properties if it is derived from sequence-based strings.[17] As far as we know, recurrent neural network (RNN) outperforms other models including CNN when dealing with sequence data, such as natural language processing[18] and electrocardiogram classification[19-21]. Even though CNN performs well due to its ability of local feature selection, RNN has its advantage in global feature discovery.[22,23] Regarding to molecules, the local feature represents the type of atoms and functional group, and global feature represents the configuration of them. Molecular properties and activities depend on not only atom types and functional groups (local features), but also the configuration of them (global features). However, very few works have been done to study the neural network performance on both local the global feature extraction for molecules.

On the other hand, scarce availability of labelled data is the major obstacle for QSAR modelling.[1] According to the probably approximately correct theory, the size of training data plays a key role in the performance of machine learning methods.[24] Nonetheless, the available data set are small at most stages of QSAR pipeline, especially biological activity modelling. One strategy to solve this problem is transfer learning algorithms.[25] Transfer learning is one kind of machine learning methods which takes advantage of existing, generalizable knowledge from other sources.[26,27] For example, Li et al. showed a RNN model pretrained on one million unlabelled molecules from ChEMBL and fine-tuned with Lipophilicity, HIV and FreeSolv data. It indicated that transfer learning improved the performance strongly compared with learning from scratch. Additionally, Iovanac et al. improve the QSAR prediction ability by the integration of experimentally available pKa data and DFT-based characterizations of the (de)protonation free energy.[29] However, few transfer learning strategies for multiple biological activity datasets have been presented. In comparison with physicochemical properties, the measurements of molecular biological activities are more time and resource consuming. Therefore, it is advisable to train more transfer learning models among this type of datasets. In this way, the information from large ligand-target dataset would be transferred into small one to facilitate molecular design and drug discovery.

Herein, we describe the convolutional recurrent neural network and transfer learning (CRNNTL) method to tackle the two problems above. The convolutional recurrent neural network (CRNN) is loosely inspired by the architectures proposed in the applications of polyphonic sound detection and electrocardiogram classification, which integrates the advantages from CNN and RNN at the same time.[30-32] Firstly, the CRNN model is tested using diverse benchmark datasets including regression and classification tasks, and compared with CNN and classical method, such as random forest (RF) and SVM. Then, a isomers-based dataset is trained by CRNN and CNN, to elucidate the ability of CRNN for both local and global feature learning. Next, we shown that the transfer learning part of CRNNTL could be used to improve the performance for scarce biological activity datasets, which depends on the binding site similarity. Finally, high versatility of CRNNTL was shown by QSAR modelling based on two other latent representation derived from different type of AEs.

**Materials and Methods**

**Input:** While latent representation could be generated from various initial molecular representation by encoders of AEs, in this work we concentrated on the sequence-based SMILES as the input representations. The SMILES represent molecular structures where atoms are labelled as nodes and bonds between atoms are encoded as edges [33]. In the beginning, we used canonical SMILES to generate latent representation. Then, data augmentation was performed to optimize our modelling in which 9 other SMILES were randomly generated from the canonical SMILES for model training.

**AEs:** Subsequently, the SMILES strings were encoded by AEs to generate fixed-length vectors as the latent representations. In recent years, developing AEs for molecular de novo design and QSAR modelling has become a hot topic for drug discovery.[34-38] In an AE, the latent representation is derived from the encoder. Next, the decoder part of an AE could be used to reconstruct the molecular structure. For instance, VAEs is one kind of developed AEs in which new samples could be generated by the decoders. Additionally, adversarial autoencoders (AAEs) are modification of VAEs where an AE is combined with a generative adversarial network (GAN). Due to a prior distribution of the training, AAEs facilitate the generation of novel structures. In this study, two VAEs[5,7] and one AAE[39] were used for latent representation generation. Their architectures are provided in Supporting information S1. At first, based on the latent representation from CDDD (one VAE algorithm)[5], we compare CRNNTL performance on the QSAR modelling and transfer learning with state-of-the-art methods. Then, the versatility was studied based on the latent representation generated from the other VAE and AAE method.

**Datasets and preprocessing:** The QSAR datasets include various physicochemical or physiological properties and biological activities in which 10 for classification and 10 for regression were selected from different sources. **Table 1** summarized short information about them Datasets were obtained from DeepChem or other sources. The isomers-based dataset represents different melting point for 70 molecules. In the dataset, for some amino acids, it is difficult to get the exactly melting point because they tend to decompose before melting. In such case, the decomposition temperatures are used as labelling data instead. Each isomer couples were in the same group for cross validation. The preprocessing method is based on the following standard: pure organic molecules, molecular weight ranging from 12 to 500, less than 3 heavy atoms and the partition coefficient (logP) between -7 and 5, removing stereochemistry. were preprocessed in the same way as the pretraining dataset.[5]

**Table 1. Overview of the Datasets used in this study**

| Acronym | Description | Size | Acronym | Description | Size |
|---|---|---|---|---|---|
| Regression | | | Classification | | |
| EGFR | Epidermal growth factor inhibition[40] | 4113 | HIV | Inhibition of HIV replication[44] | 41101 |
| EAR3 | Ephrin type-A receptor 3[40] | 587 | AMES | Mutagenicity[5] | 6130 |
| AUR3 | Aurora kinase C[40] | 1001 | BACE | Human β-secretase 1 inhibitors[44] | 1483 |
| FGFR1 | Fibroblast growth factor receptor 1 | 4177 | HERG | HERG inhibition[39] | 3440 |
| MTOR | Rapamycin target protein 1[42] | 6995 | BBBP | Blood–brain barrier penetration[44] | 1879 |
| PI3 | PI3-kinase p110-gamma[43] | 2995 | BEETOX | Toxicity in honeybees[47] | 188 |
| LogS | Aqueous solubility[44] | 1144 | JAK3 | Janus kinase 3 inhibitor[48] | 868 |
| Lipo | Lipophilicity[44] | 3817 | BioDeg | Biodegradability[49] | 1698 |
| BP | Boiling point[45] | 12451 | TOX21 | In-vitro toxicity[44] | 7785 |
| MP | Melting point[46] | 283 | SIDER | Side Effect Resource[44] | 1412 |

**QSAR model:** Our new QSAR modelling method based on CRNN were benchmarked against state-of-the-art ones, including CNN and SVM applied on latent representation as well as classical machine learning on ECFPs.

As for the CNN architecture, it contains convolutional and classification (or regression) part. The local feature learning part has 3 convolutional layers. The kernel sizes of the convolution were 5, 2 and 5. And the number of the produced filters were 15, 30 and 60, respectively. Batch normalization were used after each convolutional layer. After the pooling operation, the data went through two fully connected layers as the classification (or regression) part of which the output layer consisted of two neurons for classification tasks (or one neuron for regression). The hyperparameter and architecture optimization were shown in the next section. Early stopping was performed to avoid overfitting. The SVM modelling applied on latent representation were analysed according to the previous work. Meanwhile, Random Forest (RF) were implemented in scikit-learn for the modelling based on ECFP.

**Evaluation:** The random 5-fold cross-validation were performed to compare with the performance of the aforementioned methods. The area under the receiver characteristic curve (ROC-AUC) and coefficients of determination ($r^2$) were used for the classification and regression tasks, respectively.

**Transfer learning:** Source dataset was first trained using the CRNN model by 5-fold cross validation. Then, best performance model was selected and saved for the transfer learning use. Three transfer learning methods were used here. The first one is based on the conventional method. The pretrained model was loaded. After loading target data, the convolutional part was frozen and other parts were trained. Finally, the convolutional part was unfrozen and the whole network was fine-tuned. The second way follows the same process in the beginning. However, the difference was that there is no unfreezing for the convolutional part. In other words, the local feature learning part wasn't tuned in the knowledge transfer process. As for the third method, the GRU part was frozen in the whole training process.

**Results and Discussions**

**Model Optimization:** The performance of the deep learning method depends on the hyperparameters and network architecture. Firstly, hyperparameter optimization was performed by grid search. A series of models were built using different combinations of batch size for the whole neural network, as well as optimizer learning rate, activation functions for convolutional or GRU parts, respectively. The hyperparameter settings are summarized in Table 1. The model provided the best performance with 128 batch size and relu activation function for the two parts. While the best learning rates were 0.0001 and 0.0005 for convolutional and GRU, respectively.

**Table 2. Overview of the hyperparameter Settings by Grid Search.**

|  | convolutional | GRU | whole parts |
|---|---|---|---|
| batch size | - | - | [32, 64, 128] |
| activation function | [tanh, relu] | [sigmoid, relu] | - |
| learning rate | [0.001, 0.0005, 0.0001] | [0.001, 0.0005, 0.0001] | - |

Then, architecture optimization was performed on both convolutional and GRU parts. As for convolutional part, the model performed well with 3 convolutional layers. As the number of layers increase, little improvement or slight decrease (less than 2%) occurred as shown in Supporting information S2. This implied that deeper CNN network architecture involving more parameters would affect the final performance of the model, especially for relatively small dataset in our cases.[10] Considering the increase of training time and computing cost, three convolutional layers was suitable for the modelling. Regarding to GRU part, one bidirectional layer for the GRU was enough for model training. One of the reasons is similar with the one in CNN case. Additionally, since the local feature has been extracted well by CNN, denser GRU layers is not needed any more.

**QSAR on various datasets:** After model optimization, the QSAR results by five models were compared with each other. **Table 3** shows the results for regression datasets, while **Table 4** demonstrates the results for classification tasks. Except for a few tasks, the CRNN model with augmentation method (AugCRNN) yielded better results than any other ones applied on latent representation, as well as classical machine learning method on ECFPs. In comparison to CRNN method, the augmentation can overcome the limitation of small data size, which was found in artificial intelligence area for not only drug discovery but also other applications like computer vision and natural language procession.[50,51]

In addition, it should be noted that the CRNN provided higher ROC-AUC or $r^2$ in almost 20 tasks than CNN method (except for mp and bace), without data augmentation at the same time. As we mentioned above, the CRNN and CNN models have the same convolutional part and fully-connected structures. Therefore, it is suggested that the GRU in CRNN results in better performance in QSAR. In the research area of electrocardiogram (ECG) analysis, the local feature represents different kinds of amplitudes and intervals in short time, while global one represents the

**Table 3. Coefficient of determination (r²) for classification datasets**

| Dataset | CNN | CRNN | AugCRNN | SVM | RF[a] |
|---|---|---|---|---|---|
| EGFR | 0.67 | 0.70 | **0.71** | 0.70 | 0.69 |
| EAR3 | 0.64 | 0.68 | **0.70** | 0.65 | 0.53 |
| AUR3 | 0.55 | 0.57 | **0.61** | 0.60 | 0.54 |
| FGFR1 | 0.63 | 0.68 | **0.72** | 0.71 | 0.68 |
| MTOR | 0.64 | 0.68 | 0.69 | **0.70** | 0.66 |
| PI3 | 0.43 | 0.47 | 0.50 | **0.52** | 0.45 |
| LogS | 0.91 | 0.92 | **0.93** | 0.92 | 0.90 |
| Lipo | 0.63 | 0.67 | 0.70 | **0.73** | 0.66 |
| BP | 0.95 | 0.96 | **0.97** | 0.96 | 0.93 |
| MP | 0.47 | 0.46 | **0.52** | 0.46 | 0.45 |

The standard mean errors were shown in Supporting Information S4. Bold texts represent the best performance.

[a] Calculated with ECFP representation

**Table 4. The area under the receiver characteristic curve (ROC-AUC) for classification datasets**

| Dataset | CNN | CRNN | AugCRNN | SVM | RF[a] |
|---|---|---|---|---|---|
| HIV | 0.80 | 0.82 | **0.83** | 0.76 | 0.78 |
| AMES | 0.86 | 0.87 | 0.88 | **0.89** | **0.89** |
| BACE | 0.89 | 0.88 | 0.90 | 0.90 | **0.91** |
| HERG | 0.83 | 0.84 | **0.86** | **0.86** | 0.85 |
| BBBP | 0.88 | 0.89 | 0.91 | **0.93** | 0.89 |
| BEETOX | 0.89 | 0.91 | **0.92** | **0.92** | 0.90 |
| JAK3 | 0.72 | 0.74 | **0.77** | 0.76 | 0.76 |
| BioDeg | 0.75 | 0.77 | **0.78** | 0.74 | 0.73 |
| TOX21 | 0.75 | 0.77 | **0.78** | 0.74 | 0.73 |
| SIDER | 0.68 | 0.70 | **0.72** | 0.70 | 0.68 |

The standard mean errors were shown in Supporting Information S3. Bold texts represent the best performance.

[a] Calculated with ECFP representation

permutation and combination of them.[31] It is reported that CRNN model can learn not only amplitudes and intervals, but also their permutation and combination.[30,31] Given the good QSAR performance of our model, we hypothesize that CRNN can effectively extract all the features of molecules in which the local feature represents the type of atoms and functional group, and global feature represents the configuration of them.

Although three different learning algorithms were performed on each task, the modelling performance based on the ECFPs could be further improved by choosing best flavour of fingerprint representation. Then, it would take considerable training time due to 20 tasks. To sum up, in spite of our harsh evaluation scheme, it still indicated that CRNN can compete or outperform other baseline methods due to both the local and global feature extraction abilities.

**CRNN performance on isomers-based dataset:** So far, the difference between local and global feature of molecules has been demonstrated. It is well-acknowledged that CNN could extract local feature effectively in molecular modelling. However, little work was done for the molecular global feature extraction. One question in mind is that which model will perform best on dataset with only global feature variation. In chemistry, the most coincident case is the isomers between which the molecules have same atoms or functional groups, but different configuration. We will now explore the melting point QSAR modelling based on isomers by CNN, CRNN and SVM applied on latent representation. The isomer information is extracted from PubChem and ChemSpyder, which includes three type of isomers alcohols and ethers, amino acids and amides, as well as carboxylic acids and esters. As shown in **Table 5**, the melting point between isomers is different with each other, while the SMILES are made up of same characters with different permutation. For instance, the melting point of amino acids is surprisingly high compared with the amides. Since there is an internal transfer of a hydrogen ion from the -COOH group to the -NH$_2$ group to leave an ion with both a negative charge and a positive charge. These ionic attractions take more energy to break, so the amino acids have high melting points for the size of the molecules. While this effect won't occur in amides. Two

**Table 5.** Names, SMILES, structures and melting points of isomers

| Name | SMILES | Molecular structure | Melting point (°C) |
|---|---|---|---|
| 2-Hydroxypropanamide | CC(O)C(N)=O | | 78 |
| Alanine | CC(N)C(=O)O | | 292[a] |
| 1,3-Dimethoxypropane | COCCCOC | | -82 |
| 1,5-Pentanediol | OCCCCCO | | -16 |
| Methyl benzoate | COC(=O)c1ccccc1 | | -12 |
| Phenylacetic acid | O=C(O)Cc1ccccc1 | | 77 |

[a]Alanine decomposes before melting

other types of isomers have difference in melting point due to similar reason.

The $r^2$ of the isomers-based datasets are 0.80, 0.88 and 0.85 using CNN, CRNN and SVM, respectively. In other words, the CRNN outperforms CNN and SVM on the isomers-based melting point modelling. The relatively huge different performance (10%) between CRNN and CNN strengthened our hypothesis that CRNN has excellent ability in not only local feature, but also global feature extraction. **Figure 1** demonstrated the architecture of CRNN and its abilities in two type of feature learning. At first, the convolutional part extracted the information about type of atoms and functional groups. Then the knowledge about the configuration of the atom and functional groups was learned by the GRU part. Finally, all the features come into Highway layers for classification or regression. It should be noted that in regular melting point modelling among 20 datasets above, the $r^2$ by CRNN model is lower than the one by CNN. To elucidate the contradiction, the molecule structures in the dataset above were checked. It was found that there is no isomer in the dataset. Therefore, the global feature extraction might not be the key ability for this QSAR modelling. Moreover, since CRNN has much more parameter for fitting in the training process, small dataset (283 molecules) might contribute to under-fitting for it. Therefore, these seemingly contradictory results between two datasets have been explained. Additionally, it further explained why CRNN can compete or outperform other models among the 20 datasets above. We will later show that the transfer learning performance by CRNNTL.

**Knowledge transfer by CRNNTL:** Some of 20 datasets above were used to investigate the knowledge transfer ability of our model. In details, MP, SIDER, AUR3c and PI3 were considered as target datasets, while BP, TOX21, FGFR1, MTOR and EGFR acted as source datasets. Regarding to transfer learning for physiology and physicochemical datasets, the ROC-AUC and $r^2$ results are reported in Supporting Information S5 in detail. With the big datasets acting as source target, the transfer learning performance achieved about 5% improvement in scores (for both MP and SIDER) compared to learning from scratch, which is consistent with the previous work due to the knowledge transferred from big dataset into small one.[52]

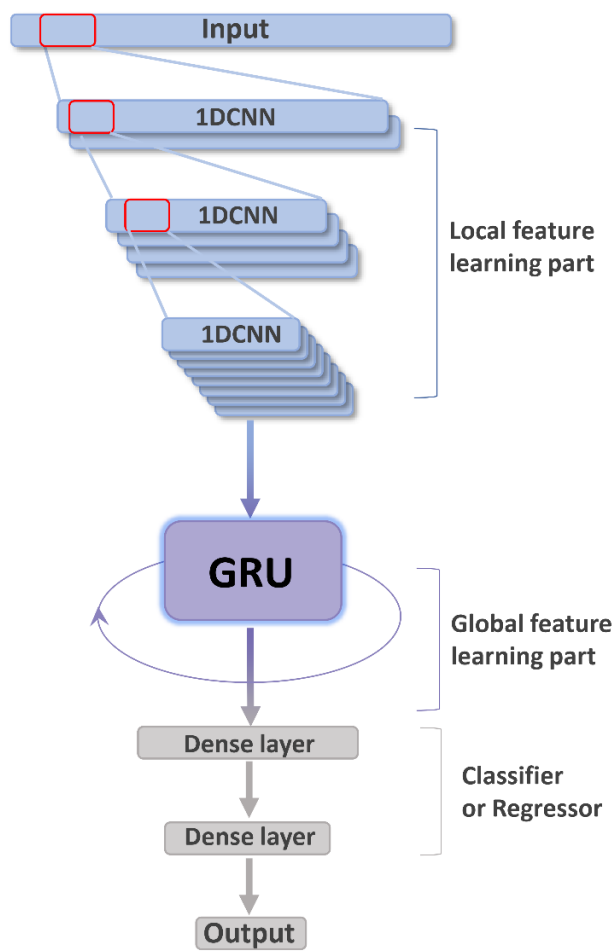

**Figure 1.** The overall architecture of the CRNN

**Table 6. Transfer learning results as PI3 and AUR3 as target datasets, and FGFR1, MOTR and EGFR as source datasets.**

| Source | PI3 | AUR3 |
|---|---|---|
| Learning from scratch | 0.47 | 0.57 |
| FGFR1 | 0.53 | 0.63 |
| MTOR | 0.61 | 0.58 |
| EGFR | 0.47 | 0.56 |

Having settled the transfer learning performance of CRNNTL in physiology and physicochemical datasets, we turn our attention to the knowledge transfer ability to biological activities. As show in the first part. The regression results of AUR3, PI3 is relatively worsen than other ones because of small data size (shown in **Table 1**). Hence, larger datasets, including FGFR1, MTOR and EGFR, were used to transfer the knowledge into AUR3 and PI3. **Table 6** showed the results of the transfer learning based on CRNN method. On the one hand, when FGFR1 serving as the source dataset, the $r^2$ of both PI3 and AUR3c increased by about 10% compared to learning from scratch. On the other hand, regarding to the MTOR, QSAR for PI3 achieved 30% improvement in $r^2$. However, no improvement was realized in case of AUR3. Despite enough performance and size of EGFR dataset, there was no statistical improvement when EGFR acting as source for both PI3 and AUR3 targets. Therefore, it demonstrated that the transfer learning ability didn't only depends on the performance and size of the source dataset.

According to lock-key model in pharmaceutical science, the protein structures play significant roles in the ligand-target interaction as well as the molecular ones, which is different from the cases in physicochemical and physiological properties. Within the target structures, local binding site similarities could be more important than global similarities.[53,54] Hence, the local binding site similarities were compared using SMAP p-values among the targets mentioned above.[55] SMAP p-values represent the binding site similarity between two targets, which is based on a sensitive and robust ligand binding site comparison algorithm.[56-58] The lower SMAP p-value, the more similarity between binding sites. As shown in **table 7**, in comparison with MTOR, SMAP p-values, 7.8e-6, is the lowest in case of PI3. Nonetheless, there is insignificant similarity between AUR3 and MTOR. These results indicated that high binding site similarity resulted in the efficient knowledge transfer from the source dataset into target dataset. When FGFR1 acting as source dataset, the results were consistent with the ones above: due to moderate binding site similarities with the source, moderate improvements were achieved for PI3 and AUR3 during the knowledge transfer process. As for EGFR, no statistical improvement in two cases arose from little binding site similarities with both two targets. In brief, strong positive correlation between transfer learning ability and binding site similarity was found.

**Table 7. Binding site similarities between different targets, SMAP p-values represent the similarities and the lower SMAP p-value, the more similarity between different targets.**

|  | PI3 | AUR3 |
|---|---|---|
| FGFR1 | 1.3e-4 | 2.1e-5 |
| MTOR | 7.8e-6 | 9.4e-3 |
| EGFR | 5.2e-3 | 8.6e-3 |

The transfer learning performance by CNN was also studied and shown in Supporting information S6. Given binding site similarities, high improvement was achieved as well. However, the result of CRNN on transfer learning was still better than CNN method. To elucidate the difference, we tried to freeze the weights either in local (convolutional) or global (GRU) feature learning part when fine-tuning the models. As shown in Supporting Information S7, similar improvement was shown when freezing the weights in local feature learning part. While there was little improvement as weights in global part frozen. It indicated that fine-tuning in GRU part played a key role in the transfer learning process. That is to say, in CRNNTL, local features can be shared between different targets with binding site similarity. Then, valuable local feature can be transferred from source dataset. Finally, modelling performance was improved efficiently by fine-tuning the global feature learning (GRU) part.

**Versatility:** To demonstrate the versatility of our strategy, two latent representations were generated from one AAE[39] and another VAE from DDC[7]. Then, their QSAR performance were studied based on different models mentioned above as well as transfer learning method. As shown in Supporting Information S8 and 9, the results were accordant with the aforementioned ones derived from CDDD VAE: firstly, CRNN outperformed other models among most datasets; secondly, the results using CRNN were better than the one by CNN in most cases due to abilities of both global and local feature extractions; thirdly, up to 30% improvements achieved by transfer learning module when taking binding site similarities into consideration. Meanwhile, the QSAR modelling performance based on the AAE wasn't as good as the results by two other VAEs. Compared with CDDD VAE, less molecules were trained in the sequence-to sequence model. Additionally, despite similar amount of training data, SMILES enumeration was used to improve the performance in another DDC encoder. Given the above, our approach has high versatility to be used as QSAR modelling and transfer learning for other latent representations from different AEs.

**Conclusion**

We demonstrate convolutional and recurrent neural network and transfer learning model (CRNNTL) for both QSAR modelling and knowledge transfer study. The CRNN part takes advantages of both convolutional and recurrent neural networks for feature extraction, as well as the data augmentation method. Experimental results on 20 datasets show the new model outperforms or competes with the state-of-art ones due to extraction ability not only on local (atom types or functional groups), but also global (molecular conformation) molecular feature. Additionally, the hypothesis above was strengthened by training and evaluating a isomers-based dataset. Even though more parameters are need to been trained without big data size, the performance of CRNN is almost 10% higher than traditional convolutional neural network due to the excellent global feature learning ability. Furthermore, separating global and local feature learning parts resulted in knowledge transfer with high efficiency from large dataset into small one. Up to 30% improvement were achieved by our transfer learning method between different biological activity datasets considering binding site similarities. Finally, CRNNTL showed high versatility by testing the model on different latent representations from other type of autoencoders. Accordingly, CRNNTL provides a new strategy to improve the performance of QSAR modelling and transfer learning. In particular, most biological activity datasets have small-data feature and include high-proportioned isomers and derivatives designed from pharmacophores. Therefore, we believe CRNNTL shows the pathway to effective knowledge transfer into small biological datasets.

## ASSOCIATED CONTENT

**Supporting Information**. The architecture of different AEs, Architecture optimization results, Standards mean errors for regression and classification, transfer learning for physiology and physicochemical datasets, transfer learning results using CNN, transfer learning using freezing process, QSAE performance using two other latent representation from different AEs

## AUTHOR INFORMATION

**Corresponding Author**

* yaqinlio809@gmail.com and xyuyig@gu.se.

**Author Contributions**

The manuscript was written through contributions of all authors.**Funding Sources**

This research was supported through grants from the European Research council.

**Notes**

The authors declare no competing financial interest.

## ACKNOWLEDGMENT

We gratefully acknowledge financial support from the European Research council (ERC-2017-StG-757733).## ABBREVIATIONS

QSAR: quantitative structure–activity relationship; ECFPs: extended-connectivity fingerprints; AEs: autoencoders; VAE: variational autoencoders; AAE: adversarial autoencoders; CNN: convolutional neural network; GRU: gated recurrent unit; SMILES: Simplified Molecular Input Line Entry Specification; SVM: support vector machine; RF: random forest; AUC-ROC: area under the receiver characteristic curve; $r^2$: coefficients of determination; CRNN: convolutional neural network; AugCRNN: CRNN with data augmentation method.

## REFERENCES

1. Le, T.; Epa, V. C.; Burden, F. R.; Winkler, D. A., Quantitative structure–property relationship modeling of diverse materials properties. *Chem. Rev.* **2012,** *112*, 2889-2919.
2. Muratov, E. N.; Bajorath, J.; Sheridan, R. P.; Tetko, I. V.; Filimonov, D.; Poroikov, V.; Oprea, T. I.; Baskin, I. I.; Varnek, A.; Roitberg, A., QSAR without borders. *Chem. Soc. Rev.* **2020,** *49*, 3525-3564.
3. Rogers, D.; Hahn, M., Extended-connectivity fingerprints. *J. Chem. Inf. Model.* **2010,** *50*, 742-754.
4. Bjerrum, E. J.; Sattarov, B., Improving chemical autoencoder latent space and molecular de novo generation diversity with heteroencoders. *Biomolecules* **2018,** *8*, 131.
5. Winter, R.; Montanari, F.; Noé, F.; Clevert, D.-A., Learning continuous and data-driven molecular descriptors by translating equivalent chemical representations. *Chem. Sci.* **2019,** *10*, 1692-1701.
6. Gómez-Bombarelli, R.; Wei, J. N.; Duvenaud, D.; Hernández-Lobato, J. M.; Sánchez-Lengeling, B.; Sheberla, D.; Aguilera-Iparraguirre, J.; Hirzel, T. D.; Adams, R. P.; Aspuru-Guzik, A., Automatic chemical design using a data-driven continuous representation of molecules. *ACS Central Sci.* **2018,** *4*, 268-276.
7. Kotsias, P.-C.; Arús-Pous, J.; Chen, H.; Engkvist, O.; Tyrchan, C.; Bjerrum, E. J., Direct steering of de novo molecular generation with descriptor conditional recurrent neural networks. *Nat. Mach. Intell.* **2020,** *2*, 254-265.
8. Winter, R.; Montanari, F.; Steffen, A.; Briem, H.; Noé, F.; Clevert, D.-A., Efficient multi-objective molecular optimization in a continuous latent space. *Chem. Sci.* **2019,** *10*, 8016-8024.
9. Popova, M.; Isayev, O.; Tropsha, A., Deep reinforcement learning for de novo drug design. *Sci. Adv.* **2018,** *4*, eaap7885.
10. Hu, S.; Chen, P.; Gu, P.; Wang, B., A deep learning-based chemical system for QSAR prediction. *IEEE J. Biomed. Health Inform.* **2020,** *24*, 3020-3028.
11. Karpov, P.; Godin, G.; Tetko, I. V., Transformer-CNN: Swiss knife for QSAR modeling and interpretation. *J. Cheminformatics* **2020,** *12*, 1-12.
12. Albawi, S.; Mohammed, T. A.; Al-Zawi, S. In *Understanding of a convolutional neural network*, 2017 International Conference on Engineering and Technology (ICET), Ieee: 2017; pp 1-6.
13. Coley, C. W.; Barzilay, R.; Green, W. H.; Jaakkola, T. S.; Jensen, K. F., Convolutional embedding of attributed molecular graphs